# Construction of a class of scalar quantum field theory models in any dimension

ROMAN GIELERAK
University of Zielona Gora
Institute of Control&Computation Engineering
rgielerak@issi.uz.zgora.pl

**Abstract**

*It is observed that certain convex envelopes of Wightman type functionals corresponding to scalar, stochastically positive quantum fields consist of Wightman type functionals only. This leads to the construction of a large classes of not quasi-free scalar quantum field theory models obeying all Wightman axioms in any dimensions.*

## 1. Introduction

The so called axiomatic quantum field [1,2,3,4] theory although has contributed significantly into our deeper understanding what really , from mathematical point of view , the notion of relativistic quantum field should be ,did not answered the question , whether in the case of d=4 dimensional space-time any example of such a field, obeying all of the Wightman axioms [1,2,3,4] and describing nontrivial scattering processes do exists. To answer the existence question like this ,among another motivations for the programme of ,the so called constructive quantum field theory has been advocated and developed [ 5,6,7,8,9,10 ]. Although , on low dimensional space-time ( d< 4 ) certain nontrivial quantum field theory models obeying all of the Wightman axioms have been constructed [ 5,6,7,8,9,10 ] this question in the case d=4 seems to be still open[ 10].

  The main motivation for the present note is to propose , in some sense , a new constructive programme , alternative to the standard constructive quantum field theory approach to the problem of constructing new models of quantum field theory obeying standard Wightman demands. Our approach is basing on the observation that the standard Wightman axioms are stable against taking a suitable convex superpositions of the Wightman type functionals on a general type ( i.e. corresponding to higher spin carrying fields and of different quantum statistics as well ) of Borchers algebras [11,12,13].In particular , starting from a suitable families of quasi-free Wightman functionals it appears that the members of the constructed convex envelopes are, in general , again Wightman type functionals and typically of non quasi-free type (in the sense that the higher orders cumulants of the constructed Wightman functions systems do not vanish for $n > 2$ .

The present note contains the corresponding results ( part of [14] ) in the case of scalar, neutral and stochastically positive quantum fields only [15,16,17,18]. The extensions to the general case is not difficult and will be presented in an another note[14], where we are dealing within Borchers algebra framework [11,12,13]for general quantum fields of Wightman type [1,2,3,4].

By the very construction our method leads to models of Wightman fields which are not obeying the so called cluster property which means that the physical vacuums in the corresponding Hilbert spaces are not unique .Certain general decompositions theorems for Wightman functionals into the pure states ( that which fulfils also the cluster decomposition property ) are well known in the context of general constructions in the Borchers algebras framework[19,20,21,22,23] and also in the case of Schwinger functional approach as well [24,25]. But the main difference and the novel aspect of our constructions presented here is that the quantum fields constructed here are kinds of statistical mixing of other, not necessarily in general ,pure Wightman functionals .

Even if we start with a family of Schwinger functionals that correspond to the ergodic ( i.e. functionals obeying the cluster property condition )with respect to the translation group action we are forming certain convex envelopes of them which have very little to common with the convex decompositions onto the pure phases labelled by the corresponding algebras of observables "living in infinity "[19,20,21,22,23,24,25].Only in a very special cases our superposition method ( see below) coincides with the corresponding decomposition into pure phases. And this is why are believing that a huge class of Wightman quantum field that can be constructed by the method presented here includes a new ! examples of Wightman quantum fields that were never appeared in the literature before. (But this does not exclude the possibility that the construction presented here was known to many researches before, however it is hardly to find exact reference confirming this point. )

The question ,whether in the constructed classes of quantum fields do exist examples which describe nontrivial scattering processes is not clear for the Author and is under investigation still [14] .However , in many particular cases we are able to prove rigorously ,using a well known triviality criterions [1,2,4] that the constructed fields are trivial one [14]. The case of d=4 , with some very interesting quantum fields construction which do not belong to the Borchers class of the free , scalar quantum field is considered in [14].Also ,several results for the case of higher -spin carrying fields and the case of many component fields is almost ready for presentation and is under preparations now[14].

Organisation of this note: in the next section we present our euclidean-time generating Schwinger functional based approach and the main results ( a version of , see [14 ]) concerning preservation of the euclidean quantum field theory axioms [18] under taking convex superpositions of such functionals is presented. Certain examples of our construction are presented in section 3 of the present note.

The list of references is far for pretending to be complete and each cited source should be read together with "and references there in " .

## 2 . Stochastically positive Schwinger functionals and theirs convex envelopes.

Let $S(R^d)$, for $d \geq 1$, stands for the space of Schwartz's type test functions topologised as usually and let $S_r(R^d)$ be its real part . The space of tempered distributions will be denoted as $S_{(r)}'(R^d)$ ,with the corresponding dualisation $<\varphi,f> = \varphi(f)$.

A functional

$$\Gamma : S(R^d) \to C \qquad (2.1)$$

will be called a Schwinger functional iff it obeys the following axioms:

**Schw ( O )**
   (i) $\qquad \Gamma(0) =1$
   (ii) $\quad$ for any f real :
           $\Gamma(-f) = \Gamma^*(f)$
*where , * corresponds to complex conjugation.*

These are normalisation and respectively neutrality condition.

**Schw (1)** (*The regularity condition* )
       *There exists a continous norm $|| \; ||$ on the space $S(R^d)$ such that for any $f \varepsilon S_r(R^d)$ the map*
           $t \varepsilon R \to \Gamma(tf)$

*can be extended to a holomorphic function of z in some circle $\{ z \in C : |z| < r_f , r_f > 0 \}$ and the following estimate hold*

$$|\Gamma(zf)| \leq \exp(|z|^e \|f\|^{e'}) \quad (2.2)$$

for some $e, e' \geq 1$

*Remark 1.* This regularity condition can be relaxed/reformulated in different aspects and directions, but for the maximal simplicity of the presentation this point will be not discussed here.

From the regularity condition imposed on the Schwinger functional (2.2) it follows that for any sequence $f_i \in S(R^d)$, $i=1:n$ the following functionals, called n-point Schwinger moments $S^n_\Gamma(f_1,\ldots,f_n)$ of $\Gamma$ do exist :

$$S^n_\Gamma(f_1,\ldots,f_n) = \frac{1}{j^n} \frac{\delta^n}{\delta_{t_1}\ldots\delta_{t_n}} (\Gamma(\sum_{i=1}^n t_i f_i))|_{t_i=0} \quad (2.3)$$

where, j stands for imaginary unit, and the following estimate is valid

$$|S^n_\Gamma(f_1,\ldots,f_n)| \leq O(n)(n!)^{\frac{1}{2}} * \prod_{i=1:n} \|f_i\| \quad (2.4)$$

The estimate (2.4) follows by the standard application of the Cauchy integral formula.

**Schw (2)** *(Reflection positivity)*
  *for any sequence of complex numbers $z_i$, any sequence of test functions $f_i$ supported on positive times $\{ x \in R^d : x_0 \geq 0 \}, i=1:n$*

$$\sum_{i,j=1:n} z_i z_j^* \Gamma(f_i - Rf_j) \geq 0 \quad (2.5)$$

*where R is time-reflection operator :*
$$(Rf)(x_0, x_1,\ldots x_{d-1}) = f(-x_0, x_1,\ldots x_{d-1}). \quad (2.6)$$

**Schw (3)** *(Stochastic positivity)*
       *for any sequence of complex numbers $z_i$, any sequence of test functions $f_i$, $i=1:n$*

$$\sum_{i,j=1:n} z_i z_j^* \Gamma(f_i - f_j) \geq 0 \quad . \quad (2.7)$$

From the nuclearity of $S'(R^d)$ [17], the Minlos Theorem [17] and the assumed Schw(3) and Schw(0) it follows that there exists an unique ,

probabilistic, cylindric set measure $d\mu_\Gamma$ on the Borel $\sigma$-algebra of sets of S'($R^d$)( a PBC measure) and such that

$$\Gamma(f) = \int e^{i\varphi(f)} d\mu_\Gamma(\varphi) \qquad (2.8)$$

In particular, then the Schwinger moments of $\Gamma$ given by (2.3) are moments of the measure $d\mu_\Gamma$

$$S^n_\Gamma(f_1,\ldots,f_n) = \int \varphi(f_1)\ldots\varphi(f_n) d\mu_\Gamma(\varphi) \qquad (2.9)$$

**Schw (4)** *(Euclidean invariance)*
*The Schwinger functional is $E(d)$ invariant, which means that for every element $(a, A)$ of the affine Euclidean group $E(d)$, where a is translation by the vector a and A stand for the (proper) rotation in $R^d$*

$$\Gamma(f_{(a,A)}) = \Gamma(f) \qquad (2.10)$$

*for any $f \in S(R^d)$, and where*
$f_{(a,A)}(x) = f(A^{-1}(x-a))$.

**Observation(1)**

Let $\Gamma$ be a Schwinger functional obeying demands Schw(0) up to Schw (4). Then the moments ($S^n_\Gamma$, n=1…)
of $\Gamma$ forms a system of tempered distributions obeying all
of the Osterwalder–Schrader axioms [18] for neutral, scalar quantum Bose field.

*Remark* 2.
In the original formulation of Osterwalder and Schrader [18] there is postulated that the corresponding system of Schwinger functions obeys the so called cluster decomposition property, which in terms of the Schwinger functional terms means that the following holds:

$$\lim_{a\to\infty} \Gamma(f + g^a) = \Gamma(f)\Gamma(g).$$

The cluster decompositon property yields the result that the reconstructed from $\Gamma$ the Wightman quantum field has an unique cyclic vaccum vector. Sometimes this situation is described that the reconstructed Wightman field is in a pure phase ( which corresponds to the statement that the corresponding to this field algebra of observables localised at infinity is trivial (which is equivalent that the translational group of Minkowski space space-like translations acts in the corresponding Hilbert space of states in an ergodic manner ).
But , in our case the constructed by the formula (2. 11 ) Schwinger functional $\Gamma_P$ ( *see below* ) is never clustering , either the measure has a support consisting of one atom , even in the case when all the functionals $\Gamma_\omega$ are clustering one.

In particular, that the standard `form of the Osterwalder-Schrader positivity condition holds [18] one use the cylindrical nature of the measure $d\mu_\Gamma$ writing the formula :

$$S^n_\Gamma(f_1,\ldots,f_n) = \int \varphi(f_1)\ldots\varphi(f_n) d\mu_\Gamma(\varphi) =$$
$$= \int_{R^n} x_1 \ldots x_n \, d\mu_{f_1 \ldots f_n}(x_1,\ldots x_n)$$

where, the finite dimensional measure $d\mu_{f_1 \ldots f_n}(x_1,\ldots x_n)$ are the cylindric projections of $d\mu_\Gamma$ onto the axises connected to directions ($\varphi$, $f_i$).From the Stone -Weierstrass Theorem we know that the algebra of functions generated by the functions of the form exp ( j n $x_i$ ) is dense in the Banach algebra $C_b(R^n)$.Using this and Schw (2) the standard reflection positivity of Osterwalder and Schrader follows by some simple approximation arguments therefore. See also the proof presented in Glimm -Jaffe monograph [5] .

*Remark 3.*
For a recent survey on the notion and applications of reflection positivity we refer to [26].

Now let ( $\Omega$ ,$\Sigma$ , dP ) be a probabilistic space and let ( $\omega \to \Gamma_\omega$) be a weakly $\Sigma$-measurable map with values in the space of regular Schwinger functionals obeying Schw ( 0) up to Schw (4) together with the uniformity of regularity which means that the continuous norms $\|\ \|$ from Schw (1) can be taken constant on the support of the measure dP.
 The main observation is the following :

 **Main observation:**
 **If ( $\Gamma_\omega$ )$_{\omega \varepsilon \Sigma}$ is a uniformly $\|\ \|$- regular, weakly $\Sigma$-measurable family of Schwinger functionals, then for any probability measure  dP on ($\Omega$ ,$\Sigma$ ) the  functional  $\Gamma_P$  defined  as :**
$\Gamma_P(f) = \int_\Omega \Gamma_\omega(f) dP(\omega)$                               (2.11)
**is again $\|\ \|$- regular Schwinger functional.**

**Proof : obvious !.**

*Remark 4.*
In the classical probability theory a convex sum of normal distributions only in a very exceptional cases is again normal distribution. The same happens here as we see in the next
section.
Let $\Gamma$ be a Schwinger functional obeying Schw(0) up to Schw (4). The cumulants of $\Gamma$ are defined

$$S^{n,\,T}{}_\Gamma\ (f_1,...,f_n) = \frac{1}{j^n} \frac{\delta^n}{\delta_{t_1}...\delta_{t_n}} log\ (\Gamma\ (\sum_{i=1}^{n} t_i f_i))_{/ti=0} \qquad (2.12)$$

As is well known the knowledge of cumulants is sufficient to restore to corresponding moments :
for , let $1_n = [1,2,...n\}$ and let then Par (n) stands for the set of all partitions ( nontrivial ) of the set of indices $1_n$. For $\pi\,\varepsilon$ Par( n) the corresponding decompositions is given by the corresponding blocks $B_\alpha$, i.e. $\pi = ( B_1,..B_k)$.

$$S^n{}_\Gamma\ (f_1,...,f_n) = \sum_{\pi \epsilon Par(n)} \prod_{\alpha=1}^{k} S_\Gamma^{|B_\alpha|,T} (f_{B_\alpha}) \qquad (2.13)$$

Also cumulants can be computed from the knowledge of the Schwinger moments :

$$S^{n,\,T}{}_\Gamma\ (f_1,...,f_n) =$$
$$\sum_{\pi \epsilon Par(n)} (|\pi|-1)!\,(-1)^{|\pi|-1} \prod_{\alpha=1}^{k} S_\Gamma^{|B_\alpha|} (f_{B_\alpha}) \qquad (2.14)$$

A Schwinger functional $\Gamma$ is called quasi-free functional iff all its cumulants $S^{n,\,T}{}_\Gamma$ for n >2 vanish.
Providing that $\Gamma$ is quasi-free functional it follows from (2.13) that the corresponding moments of it are given by the formulas :

$$S^n{}_\Gamma\ (f_1,...,f_n) = \sum_{\pi \epsilon 2Par(n)} \prod_{\alpha=1}^{k} S_\Gamma^{|B_\alpha|,T} (f_{B_\alpha}) \qquad (2.15)$$

where 2Par stands for the set of partitions consisting only of blocs of size not bigger then 2.
*Remark 6.*
Even if almost all Schwinger functionals $\Gamma_\omega$ in (2.11 ) obey the cluster decomposition property , only in a very special , trivial situations corresponding to the assumption that the support of consists of exactly one atom only as can be seen from the formula

$$\lim_{|a|\to\infty} \Gamma_P(f + g^a) =$$
$$\lim_{|a|\to\infty} \int_\Omega dP(\omega) \Gamma_\omega(f + g^a)$$
$$=^? \int_\Omega dP(\omega) \Gamma_\omega(f) \Gamma_\omega(g)$$
$$\neq \text{(in general !)} \Gamma_P(f) P(g).$$

### 3.1 Convex envelopes of massive, free scalar fields in any dimensions

From the Kallen-Lehmann[1,2] theorem we know that for any scalar Wightman field the corresponding two-point Schwinger function can be represented by the following formula:

$$S^2_\rho(f,g) = \int_0^\infty d\rho(m^2) S^2_{0,m^2}(f,g) \tag{3.1}$$

where $d\rho$ is a certain tempered, Borel measure supported on $[0,\infty)$ and

$$S^2_{0,m^2}(f,g) = \frac{1}{(2\pi)^{\frac{d}{2}}} \int_{\mathbb{R}^d} dk\, \hat{f}(-k)\hat{g}(k)(k^2 + m^2)^{(-1)} \tag{3.2}$$

and where $\hat{f}$ stands for the Fourier transform of $f$.
See, i.e. [1,2].

**Lemma 3.1**
*Let $\Gamma^0_\rho$ be Schwinger functional given by the formula*

$$\Gamma^0_\rho(f) = \exp\left(-\frac{1}{2} S^2_\rho(f,f)\right) \tag{3.3}$$

*where it is assumed (for simplicity only) that for d=2 the support of $d\rho$ is contained in the seminterval $[m,\infty)$, $m>0$.*
*Then:*
***1.** the functional $\Gamma^0_\rho$ obeys all the postulates Schw(0) up to Schw(4)*
*2 the functional $\Gamma^0_\rho$ is quasi-free and the moments of $\Gamma^0_\rho$ are given by the formula:*

$$S^n_{\Gamma^0_\rho}(f_1,...,f_n) = \sum_{\pi \in 2Par(n)} \prod_{\alpha=1}^k S^{|B_\alpha|,T}_\Gamma(f_{B_\alpha}) \tag{3.4}$$

3. *If moreover the support of dρ is bounded from below by some $m_*^2 > 0$ then the expressing regularity estimate can be taken as*

$$|\Gamma^0_\rho (zf)| \leq \exp (|z|^2 O(1) \|f\|^2_{-1}) \qquad (3.5)$$

where $\| \ \|_{-1}$ stands for the corresponding Sobolev like norm.

The corresponding quantum field structure reconstructed from the quasi-free functional $\Gamma^0_\rho$ is called generalized free field and is well known example of Wightman quantum field theory, however it is trivial theory from the point of view of scattering processes [1,2,3].

Now let dP be any probability, Borel measure supported on the semintervall $(m_*^2, \infty)$ with $m_*^2 > 0$.

**Proposition 3.1**
*Let $\Gamma^0_m$ stands for the Schwinger functional of the free, scalar, massive field, i.e.*

$$\Gamma^0_m (f) = \exp (-\tfrac{1}{2}\|f\|^2_{-1,m}). \qquad (3.6)$$

*Then, for any measure dP as above the functional*

$$\Gamma^P (f) = \int dP(m^2) \exp (-\tfrac{1}{2}\|f\|^2_{-1,m}). \qquad (3.7)$$

*gives rise the Schwinger functional obeying Schw (0) up to Schw ( 4). If moreover the support of dP consists of at least two points then the constructed Schwinger functional is **not** quasi-free functional.*

Example 3.1.
Let us take $dP(m^2) = \tfrac{1}{2}(\delta(m^2 - m_1^2) + \delta(m^2 - m_2^2))$, $m_1 \neq m_2$
Then the 2-point Schwinger function is given as
$$S_2^P(f,g) = \tfrac{1}{2} S^2_{0,m_1^2}(f,g) + \tfrac{1}{2} S^2_{0,m_2^2}(f,g) \qquad (3.8)$$
Let us compute the 4 point truncated moment:

$$S_4^{P,T}(f_1,\ldots,f_4) =$$
$$S_P^{4,T}(f_1,\ldots,f_4) = S_P^4(f_1,\ldots,f_4) - \sum_{\pi=(B_1,B_2)\varepsilon 2Par(4)} S_P^2(f_{B_1}) S_P^2(f_{B_2}) =$$

$$= \tfrac{1}{4} S^4_{0,m_1}(f_1,\ldots,f_4) + \tfrac{1}{4} S^4_{0,m_2}(f_1,\ldots,f_4) -$$
$$\tfrac{1}{4}\sum_{\pi=(B_1,B_2)\varepsilon 2Par(4)}(S^2_{0,m_1}(f_{B_1}) S^2_{0,m_2}(f_{B_2}) + S^2_{0,m_1}(f_{B_1}) S^2_{0,m_2}(f_{B_2}))$$

which is definitely nonzero, but equal to zero in the case $m_1=m_2$.
Similarly, one can see in an explicit form that the higher order cumulants of the Schwinger functional are all nonzero for $m_1 \neq m_2$.
In the classical probability theory the no vanishing of higher order cumulants for a finite sequence of random elements is expressing theirs statistical correlations. It seems that a similar remark in the context of our convex envelope constructions presented is that, also on the level of Schwinger moments our construction is almost linear but on the level of the corresponding quantum fields this corresponds to a highly nonlinear transformations of them.
The important observation is that this envelope taking operation can be iterated several times, each time leading to a new class of Wightman fields.

**Iteration of the procedure :**

*Now let ( $dP_\alpha$ ) be a weakly measurable family of measures on ( $m, \infty$ ) indexed by some Borel space ( $\Sigma, \beta(\Sigma)$ ).*

*As a **first step** in our construction we construct the corresponding Schwinger functionals $\Gamma^\alpha = \Gamma_{P_\alpha}$ taking $dP^\alpha$ and proceeding as in formula ( 2.11 )*

*.In the **second step** of our construction we take a two-point Schwinger moment $S^\alpha_2$ of the Schwinger functional $\Gamma^\alpha$ and we construct the quasi-free generalized free field $\Gamma^\alpha$ as in the formula (3.3 ).*

*And then, in the **third step** we are taking a probability measure $d\lambda(\alpha)$ on $(\Sigma, \beta(\Sigma))$ and, then we construct again ( in general case ) new class of Schwinger functionals, the class in general case **disjoint** with the class indexed by the space of measures dP. This class is given by the formula :*

$$\Gamma^{(P,\lambda)}(f) = \int_\Sigma d\lambda(\alpha) \, exp\,(-\frac{1}{2} S^2{}_{P^\alpha}(f,f)) \qquad (3.10)$$

*The two point Schwinger moment of the constructed functional is given by the formula :*

$$S_2^{(P,\lambda)}(f,g) = \int_\Sigma d\lambda(\alpha)(\int_0^\infty dP^\alpha\,(m^2) S^2_{0,m^2}(f,g)$$

$$= \int_0^\infty d(\lambda * P^\alpha)(m^2) S^2_{0,m^2}(f,g) \qquad (3.11)$$

where $d(\lambda * P^\alpha)(m^2) = \int_\Sigma d\lambda(\alpha) P^\alpha(dm^2)$.

which is in agreement with the previous construction and the Kahlen-Lehmann theorem.

However 4-point Schwinger moment of $\Gamma^{(P,\lambda)}(f)$:

$$S_4^{(P,\lambda)}(f_1,\ldots,f_4) = \int_\Sigma d\lambda(\alpha) \left(\sum_{\pi \in 2Par(4)} \prod_{\alpha=1}^2 S^2_{P^\alpha}(f_{B_\alpha})\right) \qquad (3.12)$$

is **not !** given by the formula for $d(\lambda * P^\alpha)(m^2)$ construction as described in Proposition (3.1). However, the more important observation is that the truncated 4 point Schwinger functional is not vanishing in a typical situation as one see from the following formula :

$$S_4^{(P,\lambda),T}(f_1,\ldots,f_4)$$

$$= \int_\Sigma d\lambda(\alpha) \left(\sum_{\pi \in 2Par(4)} \prod_{\alpha=1}^2 S^2_{P^\alpha}(f_{B_\alpha})\right) -$$
$$\sum_{\pi=(B_1,B_2) \varepsilon 2Par(4)} \left(\int_\Sigma d\lambda(\alpha_1) (S^2_{P_{\alpha_1}}(f_{B_1}))\right)\left(\int_\Sigma d\lambda(\alpha_2) S^2_{\alpha_2}(f_{B_2})\right)$$

*(3.13)*

It is clear that this construction can be iterated several times, each time obtaining a new set of Schwinger functionals obeying Schw (0) up to Schw (4), thus leading to new models of quantum fields.

### 3.2 Some convex envelopes of $P(\varphi)_2$, sineGordon$_2$ models.

In low dimensional case $d \leq 3$. some models of quantum fields have been constructed [5,6,7] and the corresponding Schwinger generating functional as well, In particular case d=2 there are many different constructions [5,6,7] of models obeying all Wightman axioms for the interactions corresponding to any bounded from below polynomial and in different regions of coupling, depending of the methods used: correlation inequalities, cluster expansions,…[5,6,7] .

They are called $P(\varphi)_2$ quantum fields and among them there are many of examples in which so called $\exp(\varphi(f))$ estimate is known [5,6,7]. This is the following estimate.

Let us denote as $d\mu_P(\varphi)$ the corresponding PBC measure on the space $S_r'(R^2)$. We will say that this measure obeys $\exp(\varphi(f))$-bound iff there exists a continuous on $S(R^2)$ norm $\|\ \|$ and such that

$$\left|\int_{S'_r(R^2)} exp\left(z\varphi(f)\right)d\mu_P(\varphi)\right| \leq exp(O(1)|z|\|f\|) \quad (3.14)$$

Let us define the Schwinger functional :

$$\Gamma_P(f) = \int_{S'_r(R^2)} exp\left(i\varphi(f)\right)d\mu_P(\varphi) \quad (3.15)$$

Then , the functional $\Gamma_P$ obeying ( 3.14) obey Schw(0) up to Schw( 4). The resulting Schwinger functional depends on many parameters , coefficients of the polynomial P are among them. Therefore , a plenty of the convex envelopes procedures like that from the previous subsection 3.1 can be applied . In this way we can construct a lot of a new scalar quantum fields on the two-dimensional space-time .Whether they are of any interest to be studied in details remain to be explained.

### *3.3 Constructions in d=3*

Here the only case corresponding to the super-renormalizable interactions model that was successfully constructed by several methods [5,7,9,10] is that obtained by perturbing the free field Hamiltonian by the local $-\lambda\varphi^4$ interaction. The regularity given by $exp(\varphi)$- bound is known in this theory also [5,7,9,10] ,therefore we can play the similar game as before starting
from the Schwinger functional of the $\varphi^4$-theories thus enriching significantly the class of known Wightman , scalar fields in d=3.

### 4.Concluding remarks

The most interesting question is to develop scattering content analysis of the constructed models in order to answer the crucial question whether the constructions presented leads to some ,interesting from the point of view of physics quantum field theory models describing nontrivial interactions in between the corresponding quantum particles, see [14] for preliminary remarks on this .
 Also , it should be stressed that , also on the level of the corresponding Wightman functions the construction presented is quasi-linear, however on the level of the corresponding quantum fields, the resulting by the standard GNS -type constructions quantum fields are, in general highly nonlinear functionals of the fields used for supplying the corresponding constructions.

*Acknowledgements.*

A very valuable and accurate remarks and questions connected to the previous version of these notes made by Max Duell are very acknowledged by us . Also the remark by Prof. Karl-Hermann Neeb on the reflection positivity was significant for me.